\documentclass[letterpaper, 10 pt, conference]{ieeeconf}  %

\IEEEoverridecommandlockouts                              %

\usepackage{amsmath} %
\usepackage{amssymb}  %
\usepackage{xcolor}
\usepackage{algorithm, algpseudocode}
\usepackage{graphicx}
\usepackage{siunitx}
\usepackage{url}
\usepackage{doi}
\usepackage{hyperref}

\DeclareMathOperator{\diag}{diag}

\DeclareMathOperator{\chol}{chol}

\algnewcommand{\TitleComment}[1]{\Statex \hspace{-\algorithmicindent} \(\triangleright\) #1}

\title{\LARGE \bf
Real-Time Online Learning for Model Predictive Control using a Spatio-Temporal Gaussian Process Approximation
}

\author{Lars Bartels, Amon Lahr, Andrea Carron and Melanie N. Zeilinger%
    \thanks{The authors are with the Institute for Dynamic Systems and Control, ETH Zürich, CH-8092 Zürich, Switzerland. This work is supported by the European Union's Horizon 2020 research and innovation programme, Marie~Sk\l{}odowska-Curie grant agreement No. 953348,~\mbox{ELO-X}.}
}

\begin{document}

\maketitle
\thispagestyle{empty}
\pagestyle{empty}

\begin{abstract}

Learning-based model predictive control (MPC) can enhance control performance by correcting for model inaccuracies, enabling more precise state trajectory predictions than traditional MPC.
A common approach is to model unknown residual dynamics as a Gaussian process (GP), which leverages data and also provides an estimate of the associated uncertainty. However, the high computational cost of online learning poses a major challenge for real-time GP-MPC applications.

This work presents an efficient implementation of an approximate spatio-temporal GP model, offering online learning at constant computational complexity. It is optimized for GP-MPC, where it enables improved control performance by learning more accurate system dynamics online in real-time, even for time-varying systems.
The performance of the proposed method is demonstrated by simulations and hardware experiments in the exemplary application of autonomous miniature racing.
\\
\\
Video: \url{https://youtu.be/x4qo66R2-Ds} 

\end{abstract}

\section{INTRODUCTION}

Model predictive control (MPC) is an advanced control strategy that optimizes control actions by predicting future system behavior based on a dynamic model~\cite{garcia_model_1989}. By solving an optimization problem at each time step, MPC can handle constraints and adapt to changing conditions, making it highly effective for complex and safety-critical applications.
However, the control performance of traditional MPC relies on the predictive capabilities of the underlying model of the system dynamics. For many nonlinear systems, deriving such an accurate model analytically can be impractical. This complication motivates the incorporation of data-driven models into the MPC framework~\cite{hewing_learning-based_2020} to enhance both performance and constraint satisfaction. %

Gaussian processes (GPs) are a popular choice to model the mismatch between the nominal model and the true system dynamics in learning-based control, as they offer data-efficient model adaptation and provide formal uncertainty quantification.
GP-based MPC (GP-MPC) leverages these properties to maintain high performance and safety under model mismatch and has already been successfully applied across various domains (cf. Tab. 2 in~\cite{scampicchio_gaussian_2025}), including autonomous racing~\cite{kabzan_learning-based_2019, hewing_cautious_2020}.
However, the cubic computational scaling of exact GP inference \cite{williams_gaussian_1995} remains a critical bottleneck for real-time GP-MPC. This issue is particularly acute in online learning scenarios, where the continuous accumulation of training data eventually leads to computational latencies that exceed the real-time constraints.

\begin{figure}
    \includegraphics[width=\columnwidth]{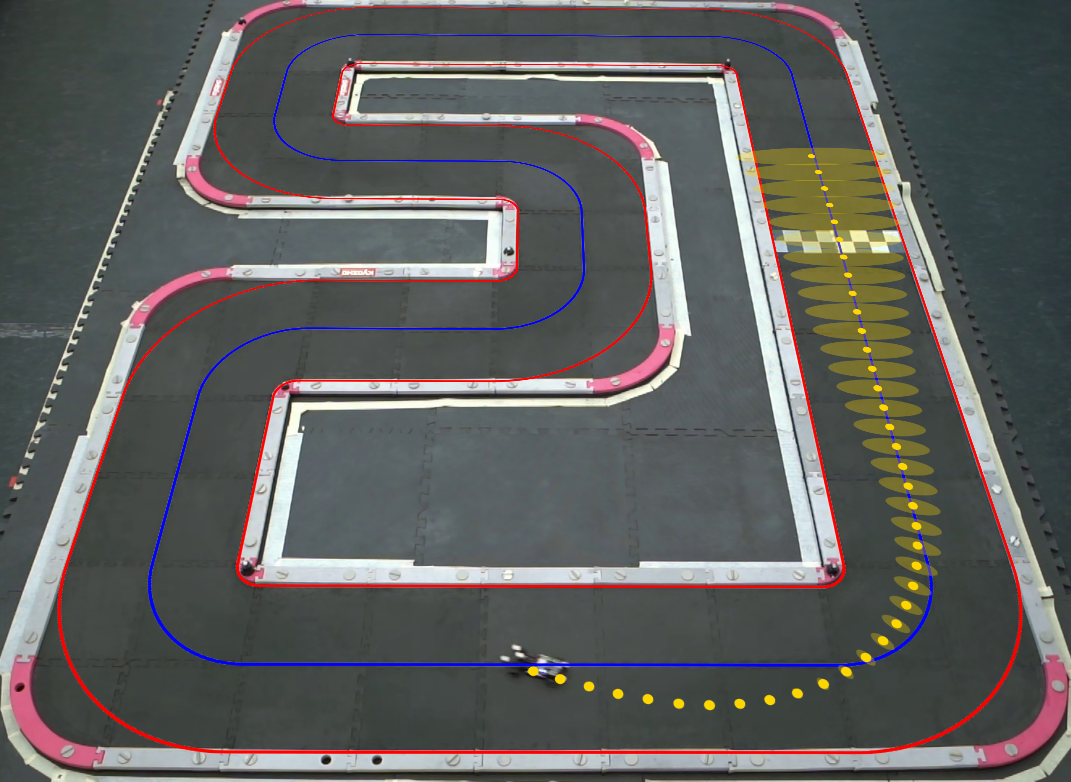}
    \caption{High-speed miniature racing despite a time-varying steering perturbation: Online learning using spatio-temporal Gaussian process approximations enables real-time adaptation to time-varying disturbances. The car aims to race along the track centerline (blue) while staying within track boundaries (red) and therefore plans its future trajectory (yellow) including uncertainty estimates based on the GP model of the residual dynamics.}
\end{figure}

\subsection{Approximate GP Inference for MPC}

To maintain real-time feasible computations,~\cite{nguyen-tuong_incremental_2011, ranganathan_online_2011, kabzan_learning-based_2019} use a \textit{subset of data} (SoD) approximation by conditioning the GP model on a \textit{data dictionary} containing a fixed number of data points selected from the full training data set. A key challenge with this approach is to decide on which data points to keep in the dictionary based on how much relevant information they contribute. 
By replacing old data points in the dictionary with newer ones,~\cite{maiworm_online_2021,bergmann_nonlinear_2022} allow the GP model to \textit{evolve}~\cite{petelin_evolving_2014}, making it suitable for dealing with time-varying disturbances.

\textit{Inducing-point} GP approximations employ a smaller set of strategically placed inducing points to summarize the training data. 
This reduces computational complexity while preserving essential information for accurate predictions~\cite{quinonero-candela_unifying_2005, titsias_variational_2009, matthews_sparse_2016}.
In the online-learning setting, 
\cite{ranganathan_online_2011,kabzan_learning-based_2019}
similarly employ a data dictionary to limit the computational complexity.
The finite-dimensional nature of inducing-point GPs also allows for
constant-time online updates by updating the inducing-point distribution~\cite{csato_sparse_2002,bijl_online_2015}.  
However, 
for time-varying models,
the incurred limited model fidelity 
necessitates to update also the inducing-point locations
to cover the extending input domain.
Variational inducing-point approximations
perform
this update 
by re-training on mini-batches of the updated training data set~\cite{hensman_gaussian_2013,cheng_incremental_2016}, 
or by performing gradient steps on the ``collapsed'' evidence lower bound~\cite{bui_streaming_2017,maddox_conditioning_2021},
which adds to 
the computation time for online inference. 
Still, to the best of the authors' knowledge, 
these methods have seen limited application in the domain of predictive control.
In this regard,
existing GP-MPC applications~\cite{hewing_cautious_2018,kabzan_learning-based_2019} have resorted to 
heuristics,
dynamically placing
inducing points 
along the previously predicted trajectory, 
since these locations represent the most relevant region in time and space at the current MPC iteration.
The dual GP approach proposed by~\cite{liu_learning_2025} instead 
employs a time-invariant inducing-point GP for long-term learning, complemented by an additional short-term model adapting to variations over time online by recursively phasing out old data.

\textit{Spatio-temporal} GPs as described in~\cite{hartikainen_kalman_2010, sarkka_spatiotemporal_2013, carron_machine_2016, todescato_efficient_2020} are naturally well-suited for application in time-varying settings.
They exploit Markovian temporal covariance structures to rewrite the GP prior as a state-space model defined through a stochastic differential equation~\cite{sarkka_applied_2019} according to their kernel, allowing to use Kalman filtering 
for learning at a constant computational cost per added data point.
In the context of MPC, a similar formulation is used by~\cite{schmid_real-time_2022}; however, it is limited to purely temporal data. The authors of~\cite{li_spatiotemporal_2023} additionally include a discrete non-temporal \textit{operating condition}, but they also do not allow for data in a continuous space as it naturally arises from a dynamic system.

To benefit from the complementary strengths of both inducing-point GPs in space and spatio-temporal GPs in time, the two approaches have been combined by~\cite{hartikainen_sparse_2011, tebbutt_combining_2021} into an approximate spatio-temporal GP model by constructing a Markov chain to track the evolution of a set of spatial inducing points over time.

\subsection{Contribution}

In this work, we develop an approximate spatio-temporal GP model based on~\cite{tebbutt_combining_2021}, tailored to the application in a GP-MPC setting.
The proposed approach enables efficient real-time inference and life-long online learning by recursively adapting to new data over time at a constant computational cost.
Unlike the recursive short-term GP of~\cite{liu_learning_2025}, it supports more intricate temporal covariances, and in contrast to other work on spatio-temporal GP-MPC~\cite{schmid_real-time_2022, li_spatiotemporal_2023}, it allows to operate over a continuous spatial input space.

We provide an open-source implementation of this model to integrate into the \texttt{L4acados} GP-MPC framework~\cite{lahr_l4acados_2024}, expanding its feature set by a GP model that is particularly suitable to online learning of a time-varying mismatch in the nominal dynamics. 
The effectiveness of the proposed method is demonstrated through experiments in the exemplary real-world application of autonomous miniature racing.

\section{PROBLEM STATEMENT}

We consider a system with the general nonlinear discrete-time dynamics
\begin{equation}
    \label{eq:fundamentals:system_dynamics}
    x(k+1) = f\left(x(k), u(k)\right) + B g\left(x(k), u(k), t_k\right) + w(k),
\end{equation}
where \mbox{$x(k) \in \mathbb{R}^{n_x}$} denotes the system state and \mbox{$u(k) \in \mathbb{R}^{n_u}$} is the control input at time $t_k$. The system dynamics are composed of the time-invariant nominal dynamics \mbox{$f: \mathbb{R}^{n_x} \times \mathbb{R}^{n_u} \to \mathbb{R}^{n_x}$}, the unknown and potentially time-varying residual dynamics \mbox{$g: \mathbb{R}^{n_x} \times \mathbb{R}^{n_u} \times \mathbb{R}_+ \to \mathbb{R}^{n_g}$} with linear mapping~\mbox{$B \in \mathbb{R}^{n_x \times n_g}$}, and random process noise~$w$ assumed as i.i.d. Gaussian with zero mean and covariance~$\Sigma_w$.

The residual dynamics are modeled as a Gaussian process such that
\begin{equation}
    \label{eq:fundamentals:residual_distribution}
    g(x, u, t) \sim \mathcal{N}\left(\mu_g(x, u, t), \Sigma_g(x, u, t)\right),
\end{equation}
where the mean $\mu_g$ and covariance matrix $\Sigma_g$ can be evaluated at the given values for $(x, u, t)$.

\subsection{Stochastic OCP Formulation}

Following the usual MPC approach, we aim to control the system with dynamics~\eqref{eq:fundamentals:system_dynamics} by optimizing state trajectories over a receding horizon of length $T$. Therefore, consider a stochastic OCP of the general form
\begin{equation}
    \label{eq:fundamentals:intractable_ocp}
    \begin{split}
        \min_{x, u} \quad & \mathbb{E} \left[ l_T(x_T) + \sum_{i=0}^{T-1} {l_i(x_i, u_i)} \right] \\
        \text{s.t.} \quad
        & x_{i+1} = f(x_i, u_i) + B g(x_i, u_i, t_i) + w_i, \\
        & \mathbb{P}(h_j(x_i, u_i) \leq 0) \geq p_j \; \text{for} \, j = 1, \dots, n_h\\
        & x_0 = x(k).
    \end{split}
\end{equation}
Here, the control task is to minimize an expected cost function defined by the sum of stage costs $l(x_i, u_i)$ and a terminal cost $l_T(x_T)$. At the same time, the system is subject to $n_h$ distinct chance constraints, imposing that the individual probability of a constraint $h_j(x_i, u_i) \leq 0$ being satisfied is at least $p_j$.

The interplay of the nonlinear nominal dynamics and the stochastic residual dynamics results in future state distributions whose exact computation is generally intractable~\cite{girard_gaussian_2002}. Hence, directly solving the stochastic OCP~\eqref{eq:fundamentals:intractable_ocp} is challenging without further simplification.
 
\subsection{Deterministic OCP Formulation}

In order to obtain a tractable OCP, we approximate the system state distribution as Gaussian, parametrized by its \mbox{mean $\mu^x$} and \mbox{covariance $\Sigma^x$} as in~\cite{hewing_cautious_2020}. Therefore, consider a first-order Taylor approximation around the mean $\mu^x$ of the uncertainty propagation 
\begin{align}
    \Sigma^x_{k+1} &= \Psi(\mu^x_k, \Sigma^x_k, u_k, t_k), \\
    \label{eq:fundamentals:uncertainty_propagation}
    \Psi(\mu^x, \Sigma^x, u, t) &= \Tilde{A} \Sigma^x \Tilde{A}^\top + B \Sigma_g(\mu^x, u, t) B^\top + \Sigma_w,
\end{align}
where $\Tilde{A} = \frac{\partial}{\partial x} [f(x, u) + B \mu_g(x, u, t)]_{x = \mu^x}$ denotes the Jacobian of the nominal and GP mean dynamics.

Using the Gaussian approximation of the state distribution, 
the stochastic chance constraints can be reformulated by deterministic tightened constraints with the tightenings given according to the uncertainty in the predictions as
\begin{equation}
    \label{eq:fundamentals:constraint_tightening}
    h_j^\Sigma(\mu^x, u, \Sigma^x) = \alpha_j \sqrt{C_j(\mu^x, u) \Sigma^x C_j^\top(\mu^x, u)},
\end{equation}
where $C_j(x, u) = \frac{\partial}{\partial x} h_j(x, u)$ and the tightening factor is given by the inverse cumulative density function of a standard Gaussian distribution as $\alpha_j = \Phi^{-1}(p_j)$.

In a common approximation to the expected cost~\cite{hewing_cautious_2020}, we consider only the mean state in the cost function, i.e.,~\mbox{$\mathbb{E}[l_i(x_i, u_i)] \approx l_i(\mu^x_i, u_i)$}.
In combination with the linearized covariance propagation~\eqref{eq:fundamentals:uncertainty_propagation} and constraint tightenings~\eqref{eq:fundamentals:constraint_tightening}, this results in a nonlinear OCP whose optimization variables~\mbox{$(\mu^x, \Sigma^x, u)$} are all defined deterministically.

\subsection{Zero-Order Optimization}

Although a deterministic OCP can be formulated with the approximations discussed above, actually solving it remains challenging due to the high number of optimization variables and computationally expensive GP evaluations. Thus, these approximations alone are often insufficient to enable a real-time capable controller.

The zero-order SQP algorithm presented in~\cite{zanelli_zero-order_2021, feng_inexact_2020, lahr_zero-order_2023} is an approach to greatly reduce computational cost and still find a feasible, albeit suboptimal, solution to the OCP.
This algorithm assumes the covariance propagation~\eqref{eq:fundamentals:uncertainty_propagation} to have zero gradient with respect to $\mu^x$ and $u$ during optimization. Following the recent implementations~\cite{frey_efficient_2023, lahr_l4acados_2024}, we further neglect the gradients of the constraint tightenings~\eqref{eq:fundamentals:constraint_tightening}.
The zero-order approximation thereby allows the elimination of the state covariances $\Sigma^x$ as optimization variables. Instead, $\Sigma^x$ and the resulting values of $h^\Sigma$ can be precomputed based on the previous iterate $(\Hat{\mu}^x, \Hat{u})$ of the optimizer.
This results in the reduced-size OCP
\begin{equation}
    \label{eq:fundamentals:zero_order_ocp}
    \begin{split}
        \min_{\mu^x, u} \; & {l_T(\mu^x_T) + \sum_{i=0}^{T-1} {l_i(\mu^x_i, u_i)}} \\
        \text{s.t.} \; 
        & \mu^x_{i+1} = f\left(\mu^x_i, u_i\right) + B \mu_g\left(\mu^x_i, u_i, t_i\right), \\
        & h_j(\mu^x_i, u_i) + \Hat{\beta}_j^{(i)} \leq 0 \; \text{for} \, j = 1, \dots, n_h, \\
        & \mu^x_0 = x(k),
    \end{split}
\end{equation}
where $\Hat{\beta}_j^{(i)} = h_j^\Sigma(\Hat{\mu}^x_i, \Hat{u}_i, \Hat{\Sigma}^x_i)$ with $\Hat{\Sigma}^x$ being computed according to~\eqref{eq:fundamentals:uncertainty_propagation} based on $(\Hat{\mu}^x, \Hat{u})$.
The algorithm~\cite{zanelli_zero-order_2021} then proposes to solve the OCP~\eqref{eq:fundamentals:zero_order_ocp} iteratively with SQP, where the fixed constraint tightenings $\Hat{\beta}$ are updated between SQP iterations.

\section{REAL-TIME SPATIO-TEMPORAL GP-MPC}

For enabling a real-time capable GP-MPC strategy with online learning, we combine the zero-order GP-MPC algorithm~\cite{lahr_zero-order_2023} with an approximate spatio-temporal GP model presented by~\cite{tebbutt_combining_2021}. On this foundation, we contribute an efficient spatio-temporal GP algorithm by specifically tailoring and optimizing the model for application in an MPC setting.

\subsection{Approximate Spatio-Temporal GP Model}

The residual dynamics~\eqref{eq:fundamentals:residual_distribution} are modeled as a GP with the spatial input features \mbox{$z = \begin{bmatrix} x^\top & u^\top \end{bmatrix}^\top$} and the temporal input feature $t$. We assume this GP to have zero prior mean for simplicity and its kernel to be separable as $k((z, t), (z', t')) = k^s(z, z') k^t(\tau)$ into a spatial kernel $k^s$ and a stationary temporal kernel $k^t$ with $\tau = t - t'$. This separability allows us to treat the spatial and temporal component of the GP independently according to the approach outlined in~\cite{tebbutt_combining_2021}.

The spatial component of the kernel is approximated using a set of inducing points $V$ placed at $M$ fixed locations in space. Let the inducing-point values $v_k \in \mathbb{R}^M$ at any fixed time $t_k$ follow a Gaussian distribution, whose mean $\mu_v^{(k)}$ and covariance $\Sigma_v^{(k)}$ may change over time.
Conditioned on $v_k$, the distribution of an observation $y_k$ at spatial inputs~$Z_k$ and time~$t_k$ only depends on the spatial kernel component $k^s$ by the GP conditional
\begin{equation}
    \label{eq:fundamentals:observation_conditional}
    y_k \mid v_k \sim \mathcal{N} \left( K_{Z_k V} K_{V V}^{-1} v_k, K_{Z_k Z_k} - Q_{Z_k Z_k} + \Sigma_\epsilon \right),
\end{equation}
where $K_{**}$ is a spatial covariance matrix such that \mbox{$[K_{AB}]_{i,j} = k^s(A_i, B_j)$} for two sets $(A,B)$ of spatial input points. Here, we define \mbox{$Q_{Z_k Z_k} = K_{Z_k Z_k} - K_{Z_k V} K_{V V}^{-1} K_{V Z_k}$} and denote by $\Sigma_\epsilon$ the measurement noise variance.
In accordance with the standard inducing-point GP inference equations~\cite{quinonero-candela_unifying_2005}, the laws of total expectation and total variance yield the distribution of the function value $g(z, t) \sim \mathcal{N}(\mu_g(z, t), \Sigma_g(z, t))$ with
\begin{equation}
    \label{eq:fundamentals:inducing_point_inference}
    \begin{split}
        \mu_g(z, t) = & K_{ZV} K_{VV}^{-1} \mu_v^{(t)}, \\
        \Sigma_g(z, t) = & K_{ZZ} - K_{ZV} K_{VV}^{-1} (K_{VV} - \Sigma_v^{(t)}) K_{VV}^{-1} K_{VZ}.
    \end{split}
\end{equation}

In the temporal component, the procedure presented in detail in~\cite{carron_machine_2016, sarkka_applied_2019} is applied. This approach uses the Fourier transform of the stationary temporal kernel by (approximately) letting $\mathcal{F}[k^t(\tau)](\omega) = W(i\omega) W(-i\omega)$ such that $W(s)$ is a rational transfer function representing a linear time-invariant state-space model with dynamics matrix $F$, input matrix $G$, and output matrix $H$. If $k^t$ is a member of the half-integer Matérn kernels, an explicit formula to compute $(F,G,H)$ is given in appendix A.
The evolution over time of the inducing points is then described by the discrete-time linear Gaussian state-space model
\begin{equation}
    \label{eq:fundamentals:state_space_model}
    \begin{split}
        \Bar{v}_0 & \sim \mathcal{N}\left(0, K_{VV} \otimes P_\infty\right), \\
        \Bar{v}_{k+1} \mid \Bar{v}_k & \sim \mathcal{N} \left( (\mathbb{I}_M \otimes A_k) \Bar{v}_k, K_{VV} \otimes Q_k \right), \\
        v_k & = (\mathbb{I}_M \otimes H) \Bar{v}_k, 
    \end{split}
\end{equation}
whose state transition matrices are defined as Kronecker products (denoted by $\otimes$) of \mbox{$A_k = \exp(\Delta t_k F)$} and \mbox{$Q_k = P_\infty - A_k P_\infty A_k^\top$}, where $\Delta t_k = t_{k+1} - t_k$, $\exp(*)$ denotes a matrix exponential, and the stationary covariance $P_\infty$ is given as the solution to the continuous-time Lyapunov equation \mbox{$F P_\infty + P_\infty F^\top = - G G^\top$}.

Since the stochastic state-space model~\eqref{eq:fundamentals:state_space_model} and the observation conditional~\eqref{eq:fundamentals:observation_conditional} are both fully linear and Gaussian, the standard Kalman filtering technique~\cite{kalman_new_1960} can be applied to recursively learn the inducing mean~$\mu_v^{(k)}$ and covariance~$\Sigma_v^{(k)}$.
For simplicity of notation, define \mbox{$\Bar{A}_{k} = \mathbb{I}_M \otimes A_k$}, \mbox{$\Bar{Q}_k = K_{VV} \otimes Q_k$}, and \mbox{$\Bar{C}_k = K_{Z_k V} K_{VV}^{-1} (\mathbb{I}_M \otimes H)$}.
The recursive estimation scheme first predicts the new mean and covariance by applying the Markovian state transition model~\eqref{eq:fundamentals:state_space_model} to the previous estimate
\begin{equation}
    \label{eq:fundamentals:kalman_prior}
    \begin{split}
        \Hat{\mu}_{\Bar{v}}^{(k)} & = \Bar{A}_{k-1} \mu_{\Bar{v}}^{(k-1)}, \\
        \Hat{\Sigma}_{\Bar{v}}^{(k)} & = \Bar{A}_{k-1} \Sigma_{\Bar{v}}^{(k-1)} \Bar{A}_{k-1}^\top + \Bar{Q}_{k-1}.
    \end{split}
\end{equation}
Then, the Kalman gain
\begin{equation}
    \label{eq:fundamentals:kalman_gain}
    K_k = \Hat{\Sigma}_{\Bar{v}}^{(k)} \Bar{C}_k^\top \left( \Bar{C}_k \Hat{\Sigma}_{\Bar{v}}^{(k)} \Bar{C}_k^\top + R_k \right)^{-1},
\end{equation}
where $R_k = K_{Z_k Z_k} - Q_{Z_k Z_k} + \Sigma_\epsilon$, allows to update the current estimate based on the most recent measurements $y_k$:
\begin{equation}
    \label{sec:theory:approx_GP:spatio_temporal_approx:kalman_filter:posterior}
    \begin{split}
        \mu_{\Bar{v}}^{(k)} & = \Hat{\mu}_{\Bar{v}}^{(k)} + K_k \left(y_k - \Bar{C}_k \Hat{\mu}_{\Bar{v}}^{(k)}\right), \\
        \Sigma_{\Bar{v}}^{(k)} & = \left(\mathbb{I}_{M d} - K_k \Bar{C}_k\right) \Hat{\Sigma}_{\Bar{v}}^{(k)}.
    \end{split}
\end{equation}

\subsection{Spatio-Temporal GP Inference for MPC}

For applicability of the spatio-temporal GP model presented above in an MPC setting, ensuring computational efficiency is of paramount importance in the algorithm design, as GP-MPC imposes strict real-time constraints on computations for GP learning and inference at each time step.

A common and powerful measure to improve efficiency is caching invariant and reoccurring expressions preemptively in order to avoid expensive repeating computations. For instance, the product~$K_{ZV} K_{VV}^{-1}$ appears multiple times in the inference equations~\eqref{eq:fundamentals:inducing_point_inference} and does not need to be recomputed.

In MPC, it usually holds that the step size $\Delta t$ between consecutive time steps is constant. This allows for additional computational savings since the discrete-time state transition matrices $A, Q$ from~\eqref{eq:fundamentals:state_space_model} in the GP model consequently also become invariant and can thus be cached. 
Also note that by choosing \mbox{$\Delta t = 0$}, we can model a time-invariant GP.

Beyond computational efficiency, prevention of numerical instabilities is another critical challenge.
Recursive algorithms can be susceptible to small errors amplifying over time and culminating in inaccurate results. 
In Kalman filtering, such errors can cause the estimated covariance matrix to become indefinite, rendering the estimation algorithm invalid. To prevent this, the proposed algorithm takes inspiration from the square-root Kalman filter~\cite{kaminski_discrete_1971} and always computes the covariance matrix in its Cholesky decomposition. While this introduces some additional computational cost, it ensures that the covariance matrix remains positive semi-definite after each update, preserving the integrity of the Kalman filter.

The proposed algorithm of an approximate spatio-temporal GP model comprises three methods, which are presented in Alg.~\ref{alg:initialize},~\ref{alg:update},~\ref{alg:evaluate} respectively. The model is first initialized by computing the state-space representation of the temporal kernel, caching invariant terms, and obtaining the prior inducing-point distribution. Then, the model can be updated to transition to the next time step and incorporate new data if it is available. Finally, the GP can be evaluated at given spatial inputs to provide an uncertainty-aware residual dynamics estimate as required for stochastic MPC.

\begin{algorithm}
    \caption{Initialize Approximate Spatio-Temporal GP}
    \label{alg:initialize}
    \begin{algorithmic}[1]
        \Require 
            \begin{itemize}
                \item[] 
                \item GP with zero mean, spatial kernel $k^s$, 
                temporal half-integer Matérn kernel $k^t$, noise $\sigma_\epsilon^2$
                \item spatial inducing-point locations $V$
                \item time step size $\Delta t$
            \end{itemize}

        \TitleComment{\textbf{state-space representation of temporal kernel}}
        \State compute $(F,G,H)$ from $k^t$ by using~\eqref{eq:appendix:state_space_matrices}
        \State obtain $P_\infty$ by solving $F P_\infty + P_\infty F^\top = - G G^\top$
        
        \TitleComment{\textbf{cache invariant expressions}}
        \State compute $K_{VV}$ by evaluating $k^s$
        \State $K_{VV}^{1/2} = \chol{(K_{VV})}$
        \State $\Bar{H} = \mathbb{I}_M \otimes H$
        \State $\Bar{A} = \mathbb{I}_M \otimes A$ with $A = \exp{(\Delta t \cdot F)}$
        \State $\Bar{Q}^{1/2} = K_{VV}^{1/2} \otimes \chol{(Q)}$ with $Q = P_\infty - A P_\infty A^\top$
        \State $K_{VV}^{-1/2} = (K_{VV}^{1/2})^{-1}$
        \State $L = K_{VV}^{-1/2} \Bar{H}$
        
        \TitleComment{\textbf{initialize inducing-point distribution}}
        \State $\mu_{\Bar{v}} \gets 0_{Md}$
        \State $\Sigma_{\Bar{v}}^{1/2} \gets K_{VV}^{1/2} \otimes \chol{(P_\infty)}$
        
        \algstore{approximate_spatio_temporal_gp}
    \end{algorithmic}
\end{algorithm}

\begin{algorithm}
    \caption{Update Approximate Spatio-Temporal GP}
    \label{alg:update}
    \begin{algorithmic}[1]
        \algrestore{approximate_spatio_temporal_gp}
        
        \Procedure{UpdateGP}{$Z \in \mathbb{R}^{n_z}$, $y \in \mathbb{R}^{n_y}$}
            \State $\mu_{\Bar{v}} \gets \Bar{A} \mu_{\Bar{v}}$
            \State $\Sigma_{\Bar{v}}^{\mathrm{root}} \gets \begin{bmatrix} \Bar{A} \Sigma_{\Bar{v}}^{1/2} & \Bar{Q}^{1/2} \end{bmatrix}$
            \If{new data $(Z,y)$ is available}
                \State compute $K_{Z Z}$ and $K_{Z V}$ by evaluating $k^s$
                \State $Q_{Z Z}^{\mathrm{root}} = K_{Z V} K_{VV}^{-1/2}$
                \State $R = K_{Z Z} - Q_{Z Z}^{\mathrm{root}} (Q_{Z Z}^{\mathrm{root}})^\top + \sigma_\epsilon^2 \mathbb{I}_{|Z|}$
                \State $\Bar{C} = Q_{Z Z}^{\mathrm{root}} L$
                
                \State $P = \Bar{C} \Sigma_{\Bar{v}}^{\mathrm{root}}$
                \State $K = \Hat{\Sigma}_{\Bar{v}}^{\mathrm{root}} P^\top \left(P P^\top + R \right)^{-1}$ \Comment{Kalman gain}
                \State $\mu_{\Bar{v}} \gets \mu_{\Bar{v}} + K \left(y - \Bar{C} \mu_{\Bar{v}}\right)$
                \State $\Sigma_{\Bar{v}}^{\mathrm{root}} \gets \begin{bmatrix} \Sigma_{\Bar{v}}^{\mathrm{root}} - K P & K \chol{(R)} \end{bmatrix}$
            \EndIf
            \State $\Sigma_{\Bar{v}}^{1/2} \gets \chol{\left( \Sigma_{\Bar{v}}^{\mathrm{root}} (\Sigma_{\Bar{v}}^{\mathrm{root}})^\top \right)}$
        \EndProcedure
        
        \algstore{approximate_spatio_temporal_gp}
    \end{algorithmic}
\end{algorithm}

\begin{algorithm}
    \caption{Evaluate Approximate Spatio-Temporal GP}
    \label{alg:evaluate}
    \begin{algorithmic}[1]
        \algrestore{approximate_spatio_temporal_gp}
        
        \Function{EvaluateGP}{$Z \in \mathbb{R}^{N \times n_z}$}
            \State compute $K_{Z Z}$ and $K_{Z V}$ by evaluating $k^s$
            \State $\Bar{C} = K_{Z V} K_{VV}^{1/2} L$
            \State $\Hat{\mu}_{\Bar{v}}^{(0)} = \mu_{\Bar{v}}$ and $\Hat{\Sigma}_{\Bar{v}}^{(0)} = \Hat{\Sigma}_{\Bar{v}}^{1/2} (\Hat{\Sigma}_{\Bar{v}}^{1/2})^\top$
            \For{$k = 1, \dots, N$}
                \State $\Hat{\mu}_{\Bar{v}}^{(k)} = \Bar{A} \Hat{\mu}_{\Bar{v}}^{(k-1)}$
                \State $\Hat{\Sigma}_{\Bar{v}}^{(k)} = \Bar{A} \Hat{\Sigma}_{\Bar{v}}^{(k-1)} \Bar{A}^\top + \Bar{Q}$
                \State $m^{[k]} = \Bar{C}_k \Hat{\mu}_{\Bar{v}}^{(k)}$ \Comment{$\Bar{C}_k = \Bar{C}^{[k,:]}$}
                \State $S^{[k]} = \Bar{C}_k \Hat{\Sigma}_{\Bar{v}}^{(k)} \Bar{C}_k^\top$
            \EndFor
            \State $\mu = \begin{bmatrix} m^{[1]} & \cdots & m^{[N]} \end{bmatrix}^\top$
            \State $\Sigma = K_{ZZ} - Q_{ZZ}^{\mathrm{root}} (Q_{ZZ}^{\mathrm{root}})^\top + \diag\left(S^{[1]}, \dots, S^{[N]}\right)$
            \State \Return $\mathcal{N}(\mu, \Sigma)$
        \EndFunction
    \end{algorithmic}
\end{algorithm}

\subsection{Python Implementation \& Integration into \texttt{L4acados}}

The proposed implementation is designed for compatibility with \texttt{GPyTorch}~\cite{gardner_gpytorch_2021}, leveraging its core functionalities for evaluating GP kernel functions and Gaussian distributions. This integration allows us to rely on the \texttt{PyTorch}~\cite{paszke_pytorch_2019} automatic differentiation feature for computing the Jacobian of GP predictions required by the optimization algorithm.

Our Python implementation of an approximate spatio-temporal GP model can be accessed as a contribution to \texttt{L4acados}\footnote{%
Code: \href{https://github.com/IntelligentControlSystems/l4acados}{https://github.com/IntelligentControlSystems/l4acados}%
}~\cite{lahr_l4acados_2024}, which provides a framework for learning-based MPC, including an efficient GP-MPC implementation utilizing the zero-order SQP optimization algorithm~\cite{lahr_zero-order_2023}.
Within \texttt{L4acados}, the proposed model can be used interchangeably with the existing standard \texttt{GPyTorch} models to capture residual dynamics as a GP. Thereby, we expand the feature set of \texttt{L4acados} by a GP model that enables GP-MPC with online learning at a constant computational cost per time step without neglecting any data.

\section{AUTONOMOUS MINIATURE RACING}

\subsection{Control Task}

To demonstrate the presented implementation, we apply a GP-MPC controller utilizing the proposed approximate spatio-temporal GP model in autonomous miniature racing simulations and on hardware\footnote{Experiment data and video material: \doi{10.3929/ethz-c-000796944}, \href{https://gitlab.ethz.ch/ics/spatio-temporal-gp-mpc}{https://gitlab.ethz.ch/ics/spatio-temporal-gp-mpc}}. The environment for these experiments is provided by the CRS platform~\cite{carron_chronos_2023}.

The car is controlled by a model predictive contouring control (MPCC) formulation as it has been presented in~\cite{liniger_optimization-based_2015}. It aims to find the optimal torque $a$ and steering angle $\delta$ to maximize progress along the track parametrized by $\theta$ while respecting the track boundaries.
For our experiments, we use an MPCC controller with a horizon length of $T = 40$ that is called at a frequency of $30 \unit{\hertz}$. Accordingly, the MPC model dynamics are discretized with a $\frac{1}{30} \unit{\second}$ time step size.

The nominal car dynamics are adopted from~\cite{carron_chronos_2023} as a dynamic bicycle model with a simplified Pacejka tire friction model~\cite{rajamani_vehicle_2011}. It has a six dimensional state containing the car's position $(x_p, y_p)$ and orientation $\varphi$ in the global reference frame as well as the associated velocities $(v_x, v_y, \omega)$ in the car's reference frame. Together with the control inputs $(a, \delta, \theta)$, which are added as an additional integrator state, this results in the system state~\mbox{$x = \begin{bmatrix} x_p & y_p & \varphi & v_x & v_y & \omega & a & \delta & \theta \end{bmatrix}^\top \in \mathbb{R}^9$} and control input~\mbox{$u = \begin{bmatrix} u_T & u_\delta & u_\theta \end{bmatrix}^\top \in \mathbb{R}^3$}.

\begin{figure}
    \centering
    \includegraphics[width=\linewidth]{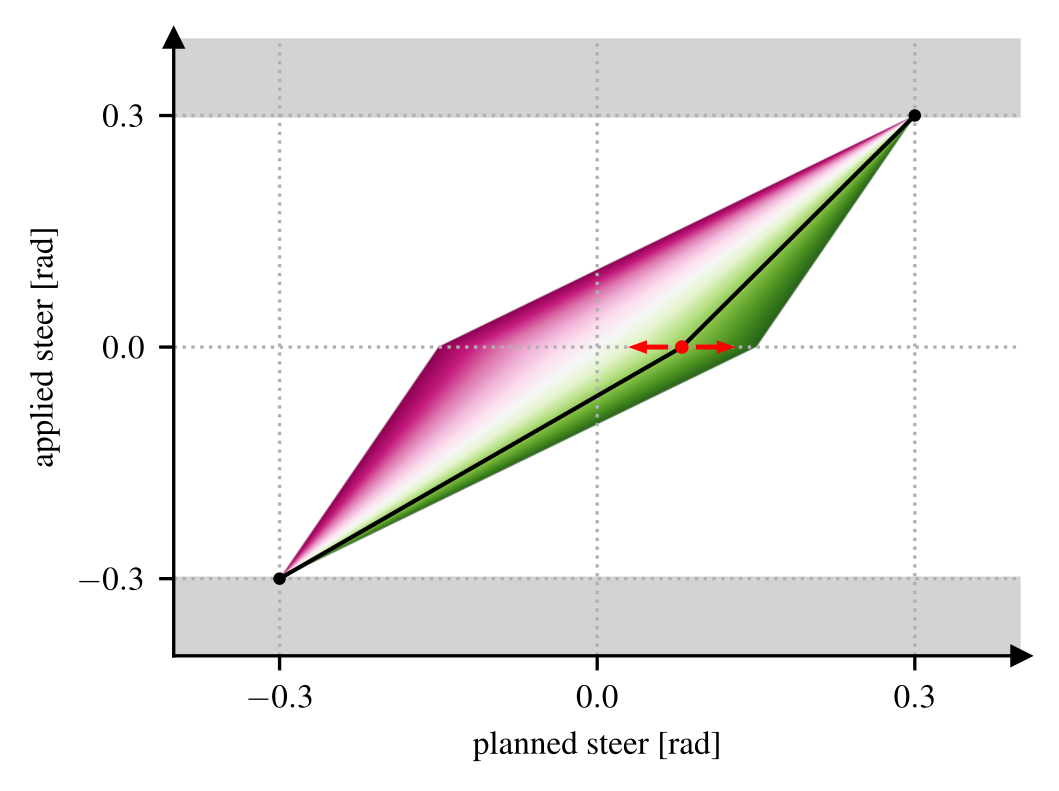}
    \caption{Steering perturbation mapping parametrized by the neutral steering offset $\delta_0$ (red dot). The shaded area illustrates the set of all employed perturbation mappings, with magenta indicating positive values and green indicating negative values of $\delta_0$.}
    \label{fig:application:perturbation}
\end{figure}

A time-varying model perturbation is introduced by dynamically altering the neutral steering offset $\delta_0$, as depicted in Fig.~\ref{fig:application:perturbation}. This setup serves as a controllable and reproducible proxy for real disturbances, which allows to isolate and evaluate the controller's online learning capability.
Note that this perturbation does not physically limit the vehicle's maneuverability. Since the full range of steering angles remains accessible, the theoretically optimal racing performance is preserved, provided the learning-based controller successfully identifies and compensates for the mismatch.
Each run begins with $\delta_0$ at its nominal value of zero, such that only inherent disturbances are present in the system. After $15 \unit{\second}$, the value of $\delta_0$ starts to vary with an amplitude of $0.15 \unit{\radian}$.

The perturbation directly affects the true dynamics of the linear and angular velocities~$(v_x, v_y, \omega)$. Hence, the residual dynamics of these three states are modeled as a GP.
The covariance function of this GP is defined as the product of a spatial RBF kernel with input features~\mbox{$(v_x, v_y, \omega, a, \delta)$} and a one-time-differentiable temporal Matérn kernel \mbox{(exponent $\nu = 1.5$)} with time $t$ as its only input feature.
For the learning-based MPCC, we then apply a zero-order GP-MPC strategy~\cite{lahr_l4acados_2024} with real-time iteration (RTI)~\cite{diehl_real-time_2005}, where the residual dynamics are modeled by the specified GP.
GP hyper-parameters are commonly obtained via minimization of the negative marginal log-likelihood or the evidence lower bound~\cite{titsias_variational_2009}, determining the parameters that best explain the training data. In our experiments, we have additionally
found that choosing a ``smoother'' GP posterior mean and covariance 
improves convergence of the optimizer, 
rendering larger lengthscale hyperparameters desirable. Consequently, we select the GP hyperparameters via gradient-based optimization of the marginal log-likelihood, incorporating lower bounds on the lengthscale values. Hyperparameter values are the same across all GPs.

A controller baseline is provided by a nominal MPCC controller, as used for example in \cite{liniger_optimization-based_2015, carron_chronos_2023}, without any awareness on the mismatch between nominal model and real system. Depending on the perturbation, open-loop predictions by the MPCC may be highly inaccurate and the car must purely rely on controller feedback to prevent crashing.
Besides the GP-MPC strategy utilizing our approximate spatio-temporal GP model implementation, we also apply an exact GP model with a spatio-temporal kernel and a conventional, purely spatial inducing-point GP model for reference. In order to maintain real-time capable online learning, the latter two rely on a subset-of-data approximation
following the GP-MPC example from \cite{lahr_l4acados_2024}.
Regardless of which model is applied, the inclusion of a GP provides the MPC controller with an estimate on the prediction uncertainty. After the first lap has been completed, online learning is activated and the GP model successively incorporates newly gathered data points, enabling it to adapt its mean and variance estimates of the residual dynamics to the real system.

\subsection{Controller Performance Comparison}

The following results were obtained from real-world racing experiments, conducted as described above.

\subsubsection{Computational Performance}
Due to the real-time constraint, we aim for the total GP-MPC computation time to remain below $30 \unit{\milli\second}$.
With the zero-order SQP RTI algorithm, only a single quadratic subproblem of~\eqref{eq:fundamentals:zero_order_ocp} is solved per time step. These computations are relatively fast (below $7 \unit{\milli\second}$), which leaves GP learning and inference as the main contributor to the total computational cost.

\begin{figure}
    \centering
    \includegraphics[width=\linewidth]{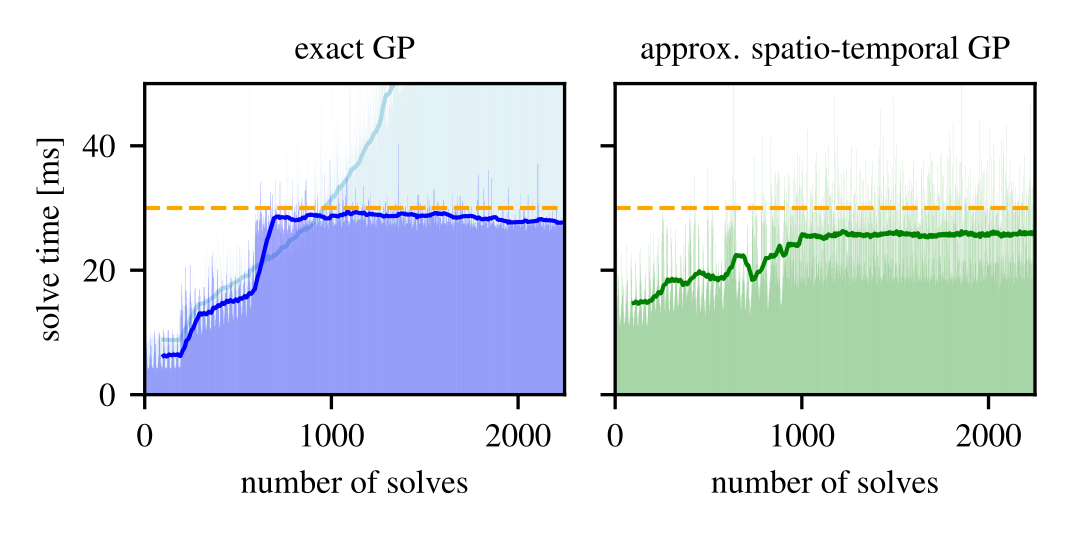}
    \caption{Experimental GP-MPC solve times using different GP models with thick lines representing a rolling average of the last 100 solves. Left: Exact GP with a SoD approximation using the 400 most recent data points (blue), compared to simulation results using the full dataset (light blue). Right: Proposed approximate spatio-temporal GP model with 80 spatial inducing points based on hardware experiments (green).}
    \label{fig:application:solve_times}
\end{figure}

The solve times plotted in Fig.~\ref{fig:application:solve_times} compare GP-MPC using our approximate spatio-temporal GP implementation with standard exact GP-MPC.
It is clear that continuously incorporating more data into an exact GP through online learning is not computationally feasible in the long term. For our experiments, we therefore used a SoD approximation, conditioning the GP only on the most recent $400$ data points and discarding the rest, which allows to mostly satisfy the computational sub-$30 \unit{\ms}$ constraint.
The approximate spatio-temporal GP is not limited through scaling with the amount of data points and it can be observed that solve times remain roughly constant at all times, allowing to run GP-MPC with online learning indefinitely without neglecting any data.

\subsubsection{Predictive Performance}
\begin{figure}
    \centering
    \includegraphics[width=\linewidth]{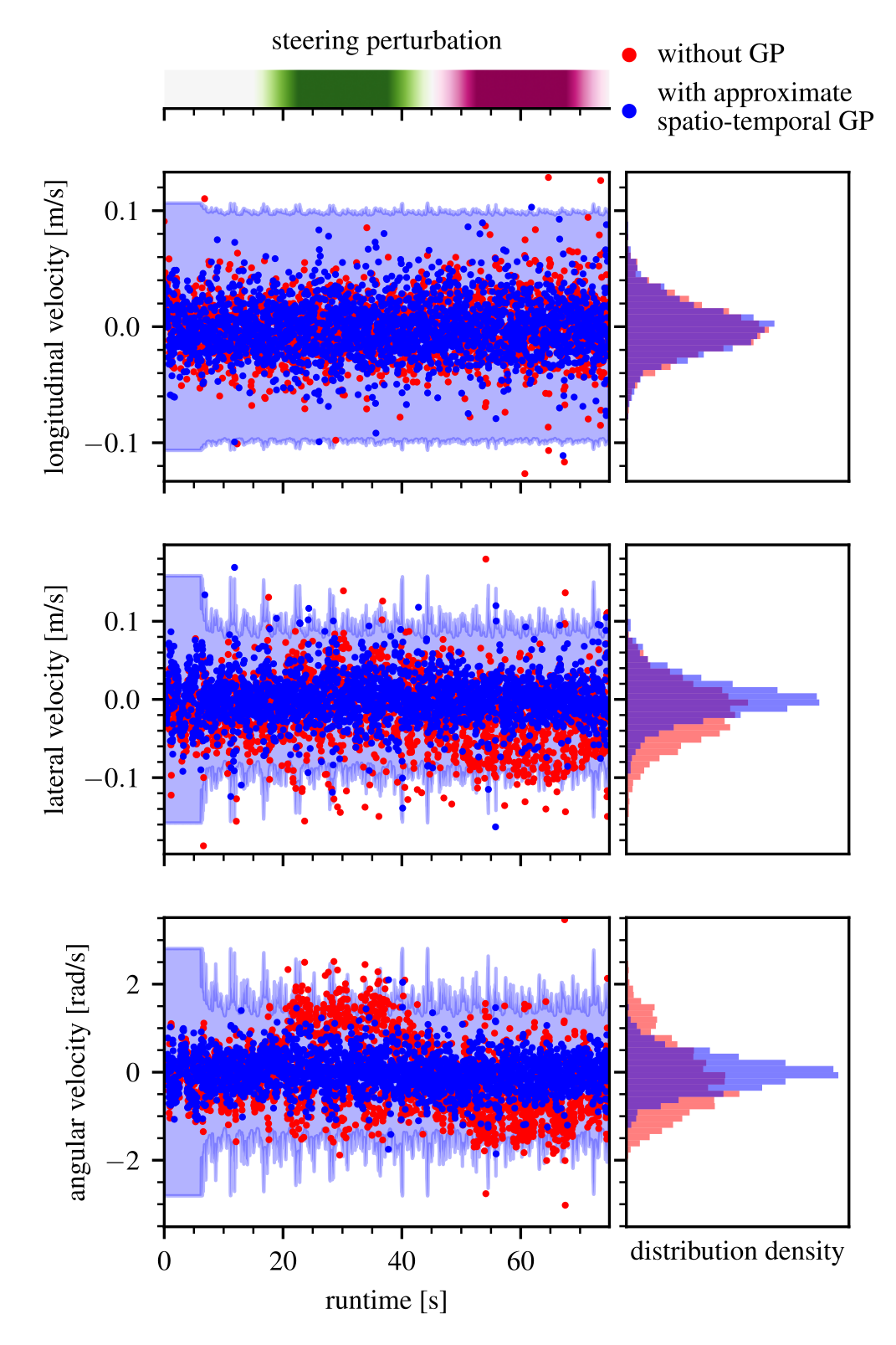}
    \caption{One-step MPC prediction error over experiment runtime without a GP (red) and with online learning by an approximate spatio-temporal GP (blue) with a \mbox{2-$\sigma$} confidence interval to model the residual dynamics. The bar up top illustrates the time-varying steering perturbation present in the system with the neutral steering offset $\delta_0$ transitioning from zero through positive (green) and negative (magenta) values and back to zero.}
    \label{fig:application:prediction_error}
\end{figure}

The efficacy of the GP’s online adaptation may be assessed by the one-step-prediction error between the car's actual state and the MPC predicted state from the previous step.
The plots in Fig.~\ref{fig:application:prediction_error} show this error for the nominal MPC and the proposed spatio-temporal GP-MPC approaches. Depending on the steering perturbation currently present in the system, it can be seen how relying on the nominal model results in a significant model mismatch in the lateral velocity and angular velocity dynamics. Learning this nominal mismatch online with the proposed approximate spatio-temporal GP model implementation considerably reduces the MPC prediction error such that the remaining residual appears independently distributed.

While not shown in the plot, the exact GP model with spatio-temporal kernel achieves a similar prediction accuracy despite the SoD truncation,
whereas the purely spatial inducing-point GP produces inconsistent prediction errors since its kernel lacks the temporal dimension necessary to distinguish between varying disturbance states at the same spatial coordinates.

\subsubsection{Racing Performance}
\begin{figure}
    \centering
    \includegraphics[width=\linewidth]{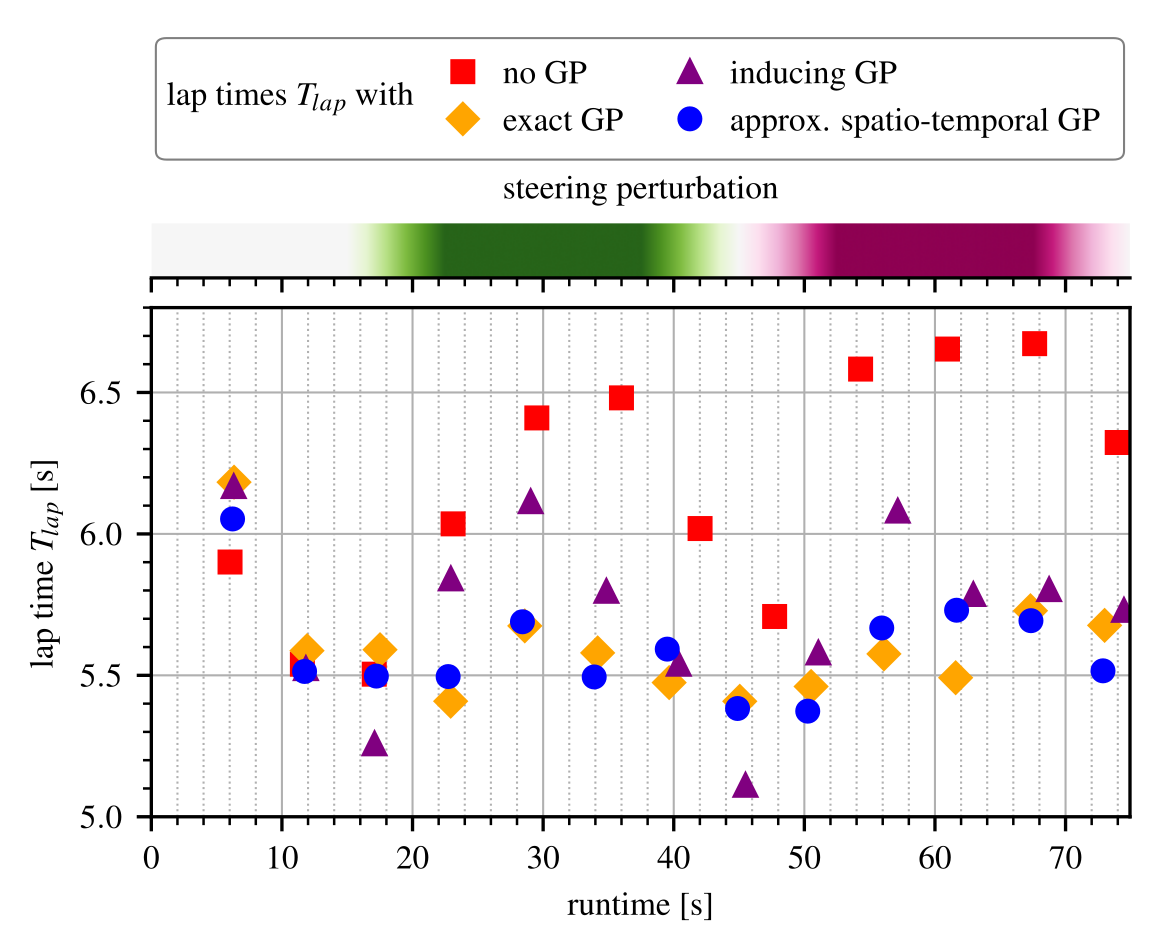}
    \caption{Evolution of the most recent lap times over the experiment runtime under a time-varying steering perturbation for nominal MPC (red squares) and GP-MPC with an exact GP model (SoD) (orange diamonds), a spatial conventional inducing-point GP model (SoD) (purple triangles), and an approximate spatio-temporal GP model (blue circles). The exact GP and spatial inducing-point GP employ a subset of data approximation and are conditioned on the most recent 400 data points. Lap times are recorded and plotted at the moment the car has completed a lap. The bar up top illustrates the time-varying steering perturbation present in the system with green corresponding to a positive neutral steering offset $\delta_0$ and magenta conversely indicating a negative value.}
    \label{fig:application:lap_times}
\end{figure}

How the reduction in prediction error affects racing performance in terms of lap times is shown in Fig.~\ref{fig:application:lap_times} for all four controller variants under the same steering perturbation.

For the first~$15 \unit{\second}$, there is no steering perturbation present and all controllers result in a similar racing performance.

As soon as the perturbation commences, it is obvious that the nominal MPC is unable to maintain this performance level.
In contrast, with the proposed approximate spatio-temporal GP model adapting to the changing conditions, the controller's performance level remains largely unaffected by the perturbation.
It can also be noted that the SoD exact GP model performs similarly well, indicating that its buffer of $400$ data points is sufficient to carry all temporally relevant information in this scenario. For the complete data persistence of our approach to offer a significant performance benefit
here, 
the residual dynamics might have to evolve more slowly relative to the data accumulation rate, causing the buffer to discard relevant data points before the system could revisit those spatial locations.

\begin{figure}
    \centering
    \includegraphics[width=\linewidth]{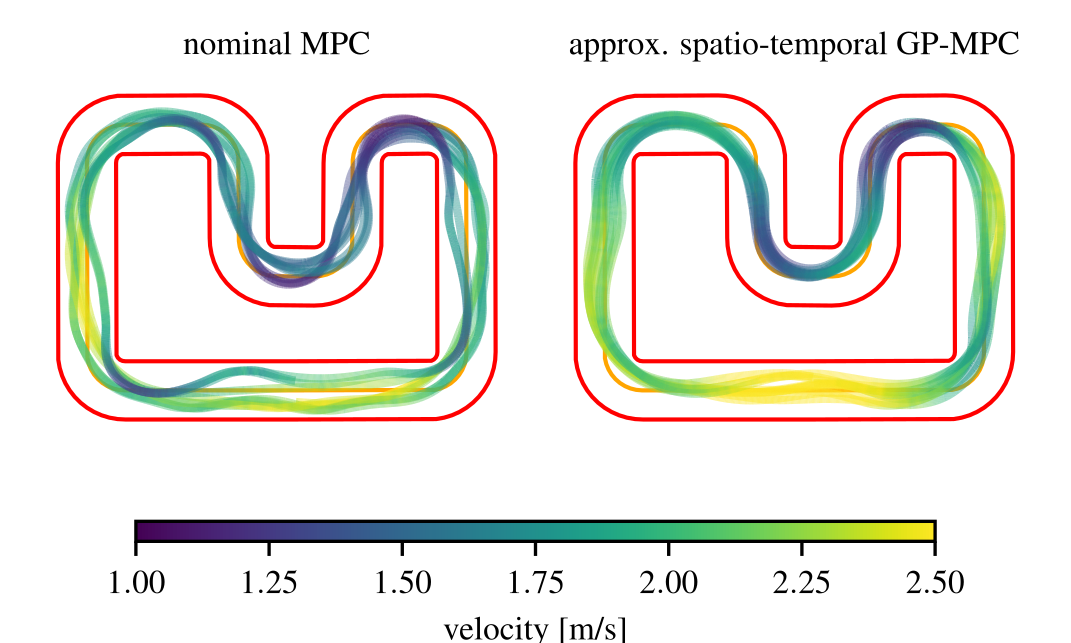}
    \caption{Trajectory of racing car on track under a time-varying steering perturbation controlled by nominal MPC without GP (left) and approximate spatio-temporal GP-MPC with online learning (right).}
    \label{fig:application:lap_trajectories}
\end{figure}

More context to the differing racing performance is given by the trajectories of the car on track over the full experiment run for nominal MPC and approximate spatio-temporal GP-MPC shown in Fig.~\ref{fig:application:lap_trajectories}.
It can be observed that the steering calibration perturbation clearly hinders the nominal controller to drive the car on a stable racing line. In particular, the car follows an undulating trajectory on the straights and the safety margin to the track boundary is violated in some corners.
Meanwhile, the GP-MPC controller is able to adapt and compensate for the changing perturbation, which allows the car to maintain much more consistent racing lines throughout the experiment runtime.

\section{CONCLUSION}

The proposed spatio-temporal GP-MPC strategy offers a principled and computationally feasible approach to predictive control with online learning. 
The underlying approximate spatio-temporal GP model preserves the flexibility and probabilistic rigor of conventional GPs, but is naturally much better suited to continuous online learning since it can be conditioned on an arbitrary number of data points without negatively affecting the computational cost of inference.
In contrast to other GP-based methods, this enables consistent real-time capability of closed-loop control without necessitating concessions such as discarding parts of the collected data. The resulting potential for performance benefits of the presented real-time implementation has been demonstrated in autonomous miniature racing experiments.
For future work, it would be interesting to investigate how hyper-parameters can be similarly updated online, and theoretical closed-loop guarantees can be established.

\section*{APPENDIX}

\subsection{State-Space Representation of Matérn Kernel}

For the stationary temporal kernel $k^t$ of the spatio-temporal GP, we focus on the family of half-integer Matérn covariance functions as they permit an exact state-space representation. The covariance between two points in time separated by $\tau$ is then given as
\begin{equation}
    \label{eq:appendix:matern_kernel}
    k^t(\tau) = \frac{2^{1-\nu}}{\Gamma(\nu)} \left(\sqrt{2\nu} \frac{\tau}{\sigma_t}\right)^\nu K_\nu\left(\sqrt{2\nu} \frac{\tau}{\sigma_t}\right),
\end{equation}
where $\Gamma$ denotes the gamma function, $K_\nu$ denotes the modified Bessel function of the second kind, $\sigma_t$ is the temporal length scale, and $\nu = D - 1/2$ for a positive integer $D$ is a smoothness parameter such that the resulting GP is $D-1$ times differentiable.

In accordance with~\cite{sarkka_applied_2019} the state-space system matrices for such a Markovian kernel can be directly computed as
\begin{equation}
    \label{eq:appendix:state_space_matrices}
    \begin{split}
        F & = \begin{bmatrix} 0 & 1 & & 0 \\ \vdots & & \ddots & \\ 0 & 0 & & 1 \\ -a_1 \gamma^D & -a_2 \gamma^{D-1} & \cdots & -a_D \gamma \end{bmatrix}, \; G = \begin{bmatrix} 0 \\ \vdots \\ 0 \\ 1 \end{bmatrix}, \\ H & = \begin{bmatrix} q & 0 & \cdots & 0 \end{bmatrix},
    \end{split}
\end{equation}
where $\gamma = \sqrt{2\nu} / \sigma_t$, $a_i = \binom{D}{i-1}$ denote binomial coefficients, and $q = \frac{((D - 1)!)^2}{(2D - 2)!} (2 \gamma)^{2D - 1}$ defines the diffusion coefficient.

\bibliographystyle{IEEEtran}
\bibliography{bibliography}

\begin{thebibliography}{10}
\providecommand{\url}[1]{#1}
\csname url@samestyle\endcsname
\providecommand{\newblock}{\relax}
\providecommand{\bibinfo}[2]{#2}
\providecommand{\BIBentrySTDinterwordspacing}{\spaceskip=0pt\relax}
\providecommand{\BIBentryALTinterwordstretchfactor}{4}
\providecommand{\BIBentryALTinterwordspacing}{\spaceskip=\fontdimen2\font plus
\BIBentryALTinterwordstretchfactor\fontdimen3\font minus
  \fontdimen4\font\relax}
\providecommand{\BIBforeignlanguage}[2]{{%
\expandafter\ifx\csname l@#1\endcsname\relax
\typeout{** WARNING: IEEEtran.bst: No hyphenation pattern has been}%
\typeout{** loaded for the language `#1'. Using the pattern for}%
\typeout{** the default language instead.}%
\else
\language=\csname l@#1\endcsname
\fi
#2}}
\providecommand{\BIBdecl}{\relax}
\BIBdecl

\bibitem{garcia_model_1989}
C.~E. Garc{\'i}a, D.~M. Prett, and M.~Morari, ``Model predictive control:
  {{Theory}} and practice---{{A}} survey,'' \emph{Automatica}, vol.~25, no.~3,
  1989.

\bibitem{hewing_learning-based_2020}
L.~Hewing, K.~P. Wabersich, M.~Menner, and M.~N. Zeilinger, ``Learning-{{Based
  Model Predictive Control}}: {{Toward Safe Learning}} in {{Control}},''
  \emph{Annu. Rev. Control Robot. Auton. Syst.}, vol.~3, no.~1, 2020.

\bibitem{scampicchio_gaussian_2025}
A.~Scampicchio, E.~Arcari, A.~Lahr, and M.~N. Zeilinger, ``Gaussian processes
  for dynamics learning in model predictive control,'' 2025.

\bibitem{kabzan_learning-based_2019}
J.~Kabzan, L.~Hewing, A.~Liniger, and M.~N. Zeilinger, ``Learning-{{Based Model
  Predictive Control}} for {{Autonomous Racing}},'' \emph{IEEE Robot. Autom.
  Lett.}, vol.~4, no.~4, 2019.

\bibitem{hewing_cautious_2020}
L.~Hewing, J.~Kabzan, and M.~N. Zeilinger, ``Cautious {{Model Predictive
  Control Using Gaussian Process Regression}},'' \emph{IEEE Transactions on
  Control Systems Technology}, vol.~28, no.~6, 2020.

\bibitem{williams_gaussian_1995}
C.~Williams and C.~Rasmussen, ``Gaussian {{Processes}} for {{Regression}},''
  \emph{Advances in neural information processing systems}, vol.~8, 1995.

\bibitem{nguyen-tuong_incremental_2011}
D.~{Nguyen-Tuong} and J.~Peters, ``Incremental online sparsification for model
  learning in real-time robot control,'' \emph{Neurocomputing}, vol.~74,
  no.~11, 2011.

\bibitem{ranganathan_online_2011}
A.~Ranganathan, M.-H. Yang, and J.~Ho, ``Online {{Sparse Gaussian Process
  Regression}} and {{Its Applications}},'' \emph{IEEE Transactions on Image
  Processing}, vol.~20, no.~2, 2011.

\bibitem{maiworm_online_2021}
M.~Maiworm, D.~Limon, and R.~Findeisen, ``Online learning-based model
  predictive control with {{Gaussian}} process models and stability
  guarantees,'' \emph{International Journal of Robust and Nonlinear Control},
  vol.~31, no.~18, 2021.

\bibitem{bergmann_nonlinear_2022}
D.~Bergmann, K.~Harder, J.~Niemeyer, and K.~Graichen, ``Nonlinear {{MPC}} of a
  {{Heavy-Duty Diesel Engine With Learning Gaussian Process Regression}},''
  \emph{IEEE Trans. Contr. Syst. Technol.}, vol.~30, no.~1, 2022.

\bibitem{petelin_evolving_2014}
D.~Petelin and J.~Kocijan, ``Evolving {{Gaussian}} process models for
  predicting chaotic time-series,'' in \emph{2014 {{IEEE Conference}} on
  {{Evolving}} and {{Adaptive Intelligent Systems}} ({{EAIS}})}, 2014.

\bibitem{quinonero-candela_unifying_2005}
J.~{Quinonero-Candela} and C.~E. Rasmussen, ``A unifying view of sparse
  approximate {{Gaussian}} process regression,'' \emph{The Journal of Machine
  Learning Research}, vol.~6, 2005.

\bibitem{titsias_variational_2009}
M.~Titsias, ``Variational {{Learning}} of {{Inducing Variables}} in {{Sparse
  Gaussian Processes}},'' in \emph{Proceedings of the {{Twelfth International
  Conference}} on {{Artificial Intelligence}} and {{Statistics}}}, ser.
  Proceedings of {{Machine Learning Research}}, vol.~5.\hskip 1em plus 0.5em
  minus 0.4em\relax Hilton Clearwater Beach Resort, Clearwater Beach, Florida
  USA: PMLR, 2009.

\bibitem{matthews_sparse_2016}
A.~G. d.~G. Matthews, J.~Hensman, R.~Turner, and Z.~Ghahramani, ``On {{Sparse
  Variational Methods}} and the {{Kullback-Leibler Divergence}} between
  {{Stochastic Processes}},'' in \emph{Proceedings of the 19th {{International
  Conference}} on {{Artificial Intelligence}} and {{Statistics}}}, ser.
  Proceedings of {{Machine Learning Research}}, vol.~51.\hskip 1em plus 0.5em
  minus 0.4em\relax Cadiz, Spain: PMLR, 2016.

\bibitem{csato_sparse_2002}
L.~Csat{\'o} and M.~Opper, ``Sparse {{On-Line Gaussian Processes}},''
  \emph{Neural Computation}, vol.~14, no.~3, 2002.

\bibitem{bijl_online_2015}
H.~Bijl, J.-W. {van Wingerden}, T.~B.~Sch{\"o}n, and M.~Verhaegen, ``Online
  sparse {{Gaussian}} process regression using {{FITC}} and {{PITC}}
  approximations,'' \emph{IFAC-PapersOnLine}, vol.~48, no.~28, 2015.

\bibitem{hensman_gaussian_2013}
J.~Hensman, N.~Fusi, and N.~D. Lawrence, ``Gaussian {{Processes}} for {{Big
  Data}},'' 2013.

\bibitem{cheng_incremental_2016}
C.-A. Cheng and B.~Boots, ``Incremental variational sparse {{Gaussian}} process
  regression,'' \emph{Advances in Neural Information Processing Systems},
  vol.~29, 2016.

\bibitem{bui_streaming_2017}
T.~D. Bui, C.~Nguyen, and R.~E. Turner, ``Streaming {{Sparse Gaussian Process
  Approximations}},'' in \emph{Advances in {{Neural Information Processing
  Systems}}}, vol.~30.\hskip 1em plus 0.5em minus 0.4em\relax Curran
  Associates, Inc., 2017.

\bibitem{maddox_conditioning_2021}
W.~J. Maddox, S.~Stanton, and A.~G. Wilson, ``Conditioning {{Sparse Variational
  Gaussian Processes}} for {{Online Decision-making}},'' in \emph{Advances in
  {{Neural Information Processing Systems}}}, vol.~34.\hskip 1em plus 0.5em
  minus 0.4em\relax Curran Associates, Inc., 2021.

\bibitem{hewing_cautious_2018}
L.~Hewing, A.~Liniger, and M.~N. Zeilinger, ``Cautious {{NMPC}} with {{Gaussian
  Process Dynamics}} for {{Autonomous Miniature Race Cars}},'' in \emph{2018
  {{European Control Conference}} ({{ECC}})}, 2018.

\bibitem{liu_learning_2025}
Y.~Liu, P.~Wang, and R.~Tóth, ``Learning for predictive control: {A} {Dual}
  {Gaussian} {Process} approach,'' \emph{Automatica}, vol. 177, p. 112316, Jul.
  2025.

\bibitem{hartikainen_kalman_2010}
J.~Hartikainen and S.~Sarkka, ``Kalman filtering and smoothing solutions to
  temporal {{Gaussian}} process regression models,'' in \emph{2010 {{IEEE
  International Workshop}} on {{Machine Learning}} for {{Signal
  Processing}}}.\hskip 1em plus 0.5em minus 0.4em\relax Kittila, Finland: IEEE,
  2010.

\bibitem{sarkka_spatiotemporal_2013}
S.~Sarkka, A.~Solin, and J.~Hartikainen, ``Spatiotemporal {{Learning}} via
  {{Infinite-Dimensional Bayesian Filtering}} and {{Smoothing}}: {{A Look}} at
  {{Gaussian Process Regression Through Kalman Filtering}},'' \emph{IEEE Signal
  Process. Mag.}, vol.~30, no.~4, 2013.

\bibitem{carron_machine_2016}
A.~Carron, M.~Todescato, R.~Carli, L.~Schenato, and G.~Pillonetto, ``Machine
  learning meets {{Kalman Filtering}},'' in \emph{2016 {{IEEE}} 55th
  {{Conference}} on {{Decision}} and {{Control}} ({{CDC}})}.\hskip 1em plus
  0.5em minus 0.4em\relax Las Vegas, NV, USA: IEEE, 2016.

\bibitem{todescato_efficient_2020}
M.~Todescato, A.~Carron, R.~Carli, G.~Pillonetto, and L.~Schenato, ``Efficient
  spatio-temporal {{Gaussian}} regression via {{Kalman}} filtering,''
  \emph{Automatica}, vol. 118, 2020.

\bibitem{sarkka_applied_2019}
S.~S{\"a}rkk{\"a} and A.~Solin, \emph{Applied {{Stochastic Differential
  Equations}}}, 1st~ed.\hskip 1em plus 0.5em minus 0.4em\relax Cambridge
  University Press, 2019.

\bibitem{schmid_real-time_2022}
N.~Schmid, J.~Gruner, H.~S. Abbas, and P.~Rostalski, ``A real-time {{GP}} based
  {{MPC}} for quadcopters with unknown disturbances,'' in \emph{2022 {{American
  Control Conference}} ({{ACC}})}, 2022.

\bibitem{li_spatiotemporal_2023}
Y.~Li, R.~Chen, and Y.~Shi, ``Spatiotemporal learning-based stochastic {{MPC}}
  with applications in aero-engine control,'' \emph{Automatica}, vol. 153,
  2023.

\bibitem{hartikainen_sparse_2011}
J.~Hartikainen, J.~Riihim{\"a}ki, and S.~S{\"a}rkk{\"a}, ``Sparse
  {{Spatio-temporal Gaussian Processes}} with {{General Likelihoods}},'' in
  \emph{Artificial {{Neural Networks}} and {{Machine Learning}} -- {{ICANN}}
  2011}.\hskip 1em plus 0.5em minus 0.4em\relax Berlin, Heidelberg: Springer,
  2011.

\bibitem{tebbutt_combining_2021}
W.~Tebbutt, A.~Solin, and R.~E. Turner, ``Combining pseudo-point and state
  space approximations for sum-separable {{Gaussian Processes}},'' in
  \emph{Proceedings of the {{Thirty-Seventh Conference}} on {{Uncertainty}} in
  {{Artificial Intelligence}}}.\hskip 1em plus 0.5em minus 0.4em\relax PMLR,
  2021.

\bibitem{lahr_l4acados_2024}
A.~Lahr, J.~N{\"a}f, K.~P. Wabersich, J.~Frey, P.~Siehl, A.~Carron, M.~Diehl,
  and M.~N. Zeilinger, ``L4acados: {{Learning-based}} models for acados,
  applied to {{Gaussian}} process-based predictive control,'' 2024.

\bibitem{girard_gaussian_2002}
A.~Girard, C.~Rasmussen, J.~Q. Candela, and R.~{Murray-Smith}, ``Gaussian
  {{Process Priors}} with {{Uncertain Inputs Application}} to {{Multiple-Step
  Ahead Time Series Forecasting}},'' in \emph{Advances in {{Neural Information
  Processing Systems}}}, vol.~15.\hskip 1em plus 0.5em minus 0.4em\relax MIT
  Press, 2002.

\bibitem{zanelli_zero-order_2021}
A.~Zanelli, J.~Frey, F.~Messerer, and M.~Diehl, ``Zero-{{Order Robust Nonlinear
  Model Predictive Control}} with {{Ellipsoidal Uncertainty Sets}},''
  \emph{IFAC-PapersOnLine}, vol.~54, no.~6, 2021.

\bibitem{feng_inexact_2020}
X.~Feng, S.~D. Cairano, and R.~Quirynen, ``Inexact {{Adjoint-based SQP
  Algorithm}} for {{Real-Time Stochastic Nonlinear MPC}},''
  \emph{IFAC-PapersOnLine}, vol.~53, no.~2, 2020.

\bibitem{lahr_zero-order_2023}
A.~Lahr, A.~Zanelli, A.~Carron, and M.~N. Zeilinger, ``Zero-order optimization
  for {{Gaussian}} process-based model predictive control,'' \emph{European
  Journal of Control}, vol.~74, 2023.

\bibitem{frey_efficient_2023}
J.~Frey, Y.~Gao, F.~Messerer, A.~Lahr, M.~Zeilinger, and M.~Diehl, ``Efficient
  {{Zero-Order Robust Optimization}} for {{Real-Time Model Predictive Control}}
  with acados,'' 2023.

\bibitem{kalman_new_1960}
R.~E. Kalman, ``A {{New Approach}} to {{Linear Filtering}} and {{Prediction
  Problems}},'' \emph{Journal of Basic Engineering}, vol.~82, no.~1, 1960.

\bibitem{kaminski_discrete_1971}
P.~Kaminski, A.~Bryson, and S.~Schmidt, ``Discrete square root filtering: {{A}}
  survey of current techniques,'' \emph{IEEE Transactions on Automatic
  Control}, vol.~16, no.~6, 1971.

\bibitem{gardner_gpytorch_2021}
J.~R. Gardner, G.~Pleiss, D.~Bindel, K.~Q. Weinberger, and A.~G. Wilson,
  ``{{GPyTorch}}: {{Blackbox Matrix-Matrix Gaussian Process Inference}} with
  {{GPU Acceleration}},'' 2021.

\bibitem{paszke_pytorch_2019}
A.~Paszke, S.~Gross, F.~Massa, A.~Lerer, J.~Bradbury, G.~Chanan, T.~Killeen,
  Z.~Lin, N.~Gimelshein, L.~Antiga, A.~Desmaison, A.~Kopf, E.~Yang, Z.~DeVito,
  M.~Raison, A.~Tejani, S.~Chilamkurthy, B.~Steiner, L.~Fang, J.~Bai, and
  S.~Chintala, ``{{PyTorch}}: {{An Imperative Style}}, {{High-Performance Deep
  Learning Library}},'' in \emph{Advances in {{Neural Information Processing
  Systems}}}, vol.~32.\hskip 1em plus 0.5em minus 0.4em\relax Curran
  Associates, Inc., 2019.

\bibitem{carron_chronos_2023}
A.~Carron, S.~Bodmer, L.~Vogel, R.~Zurbr{\"u}gg, D.~Helm, R.~Rickenbach,
  S.~Muntwiler, J.~Sieber, and M.~N. Zeilinger, ``Chronos and {{CRS}}:
  {{Design}} of a miniature car-like robot and a software framework for single
  and multi-agent robotics and control,'' in \emph{2023 {{IEEE International
  Conference}} on {{Robotics}} and {{Automation}} ({{ICRA}})}, 2023.

\bibitem{liniger_optimization-based_2015}
A.~Liniger, A.~Domahidi, and M.~Morari, ``Optimization-based autonomous racing
  of 1:43 scale {{RC}} cars,'' \emph{Optimal Control Applications and Methods},
  vol.~36, no.~5, 2015.

\bibitem{rajamani_vehicle_2011}
R.~Rajamani, \emph{Vehicle {{Dynamics}} and {{Control}}}.\hskip 1em plus 0.5em
  minus 0.4em\relax Springer Science \& Business Media, 2011.

\bibitem{diehl_real-time_2005}
M.~Diehl, H.~Bock, and J.~Schloder, ``Real-{{Time Iterations}} for {{Nonlinear
  Optimal Feedback Control}},'' in \emph{Proceedings of the 44th {{IEEE
  Conference}} on {{Decision}} and {{Control}}}, 2005.

\end{thebibliography}

\end{document}